\documentclass[12pt, a4paper]{article}

\title{Modeling of Volatility with Non-linear Time Series Model}
\author{\textrm{\textsuperscript{a}Kim Song Yon, Kim Mun Chol} $~~$ $~~$\\  \\
                  {\small{Faculty of Mathematics, \textbf{Kim Il Sung} University, Pyongyang, D. P. R. Korea}}\\
		  {\small{\textsuperscript{a}Corresponding author. e-mail address: songyonkim@yahoo.com}}
}
    
\date{}  	
\usepackage{latexsym}   
\usepackage[pdftex]{graphicx}
\usepackage{amsmath, amsthm, amsfonts, amssymb}
\usepackage{mathrsfs}
\usepackage{hyperref}
\usepackage{times}	

\usepackage{CJKutf8}

 \newtheorem{thr}{Theorem}

\begin{document}
\maketitle

%
 % --------------  abstract  -----------------
 %	
\begin{abstract}
In this paper, non-linear time series models are used to describe volatility in financial time series data. 
To describe volatility, two of the non-linear time series are combined into form TAR (Threshold Auto-Regressive Model) with 
AARCH (Asymmetric Auto-Regressive Conditional Heteroskedasticity) error term and its parameter estimation is studied.\\
\end{abstract}
{\bf Keywords:} Volatility, Non-linear time series model, ARCH, AARCH (Asymmetric ARCH), TAR, QMLE (Quasi Maximum Likelihood Estimation)

%
%
% --------------  1  Introduction  -----------------
%
%	
\section{Introduction}
In financial practice, the volatility of asset price is an important variable and its modeling has a great significance in investment, 
monetary policy making, financial risk management and etc. 

A good forecast of the volatility of asset prices over the investment holding period is a good starting point for assessing investment risk. 
The volatility of asset is the most important variable in the pricing of derivative securities. 
To price an option we need to know the volatility of the underlying asset from now until the option expires. 
In fact, the market convention is to list option prices in terms of volatility units. 
Nowadays, the definition and measurement of volatility may be clearly specified in the derivative contracts. 
In these new contracts, the volatility now becomes the underlying “asset”.

Volatility model has become one of the major tasks in analysis of financial time series models and interested many scientists.

In 1973, the mathematical finance entered a new phase by Black and Scholes \cite{bla}. 
They used a stochastic analysis theory to estimate the price of an option, the holder of which has the right to buy a good (or {\it underlying asset}) with a predetermined price $K$ at a predetermined future time $T+\tau$ rather than the present time $T$. 
(This kind of option is called an European option.)

Assume that the dynamics of the price $S_t$ of the underlying asset follows 
\begin{equation*}
dS_t = rS_tdt + \sigma S_td\omega_t
\end{equation*}
under the risk neutral measure. Here $r$ (short rate) and $\sigma$ are constants and $\omega_t$ is a standard Wiener process. 
Then the time $t$-price of the option is provided as follows \cite{bla}:
\begin{equation*}
C_t = S_t\Phi(d) - K \exp(-rt)\Phi(d-\sigma\sqrt{\tau}).
\end{equation*}
Here $\tau = T-t$  and the parameter $\sigma$ is only unknown. 

This $\sigma$ is called the {\it volatility} and as shown in the above formula, accurate estimation of $\sigma$ has become a 
very important issue in option pricing and estimating. 
Furthermore, other issues such as estimation of volatility $\sigma t$ related to the time $t$ have been raised. 
     
In 1982, Robert Engle suggested a new model which can estimate the volatility in a more accurate way \cite{eng}. 
He paid attention to the error term in ARCH, which was mostly ignored in linear time series models such as AR, ARMA 
and ARIMA, and proposed a new nonlinear model by adding, instead of simple white noise, the error-term characteristic heteroskedasticity in which the conditional deviation changes auto-regressively. 
He proved that if the stock price $X_t$ is to be changed into the following 
\begin{equation*}
\varepsilon_t = 100\log(x_t/x_t-1) ~~ \textnormal{or} ~~ \varepsilon_t = \frac{x_t-x_{t-1}}{x_{t-1}},
\end{equation*}   
then, it is possible to be modeled as follows:
\begin{equation*}
\left\{ 
\begin{array}{l}
		\varepsilon_t = z_t h_t^{\frac{1}{2}}, \\
		z_t : i.i.d, N(0, 1), \\
		h_t = \alpha_0 + \alpha_1 \varepsilon_{t-1}^2 + \cdots +\alpha_q \varepsilon_{t-q}^2.
\end{array} \right.
\end{equation*} 
This is the famous ARCH (q) model. ($h_t$ being the volatility.) 
This model overwhelmed the statisticians and (in particular) econometricians and became instantly popular. 
Engle was awarded the Nobel Prize for Economics in 2003.  
These types of non-linear models have been recognized as far superior estimation tools and meanwhile researches have been made to improve the model for better applicability. 
As a result, many modifications have appeared to form an ARCH family.  
Since then, other researches have been actively carried out to estimate volatility with ARCH (q) model. 
Engle also proposed the method of testing the ARCH effect, the effect of conditional deviation and proved the existence of Auto-Regressive Conditional Heteroskedasticity, 
therefore raising questions towards I.I.D (Independent Identical Distribution) used for the error-term previously.

In 1986, Bollerslev modified Engle's ARCH (q) model into GARCH (p, q) model \cite{bol}.
\begin{equation*}
\left\{ 
\begin{array}{l}
		\varepsilon_t = z_t h_t^{\frac{1}{2}}, z_t : i.i.d, \\
		h_t = \alpha_0 + \sum_{i=1}^q \alpha_i \varepsilon_{t-i}^2 + \sum_{i=1}^p \beta_i h_{t-i}.
\end{array} \right.
\end{equation*}  
In his paper, he proposed existence, stationary condition and MLE (maximum likelihood estimation) of the GARCH (1, 1) process. 
After this, many ARCH models such ARCH-M, IGARCH and LogGARCH have been developed.
Throughout the researches, the volatility has been proved to be more influenced by ‘bad news’ rather than ‘good news’, that is, to be asymmetric, which resulted in researches on asymmetric models.

In 1991, Nelson \cite{nel} proposed an exponential GARCH (EGARCH) model which shows the asymmetric shock.
\begin{eqnarray*}
&& h_t = \gamma_0 + \gamma_1 h_t-1+g(\varepsilon_t-1), \\
&& g(x) =\omega x + \lambda (|x|-E|x|).
\end{eqnarray*} 

But in many research papers, the effective parameter estimation and stationary conditions was not being clearly explained and this difficulty was thought to be hard to overcome \cite{har}.
But in 1993, Glosten tried to model asymmetric volatility using Threshold ARCH (TARCH) model and later, many asymmetric models were proposed \cite{glo}.

Especially in 2003, Liu Ji Chun developed the asymmetric ARCH (q) model \cite{liu}.
 
Until now, the constant researches are being made to come up with better volatility models which can show effects of various ARCH models.

In this paper, the modeling of volatility with the non-linear time series model is observed based on the analysis of previous researches. 

It is already well known that volatility and other financial time-series data are well described by ARCH models. 
At the same time, these data have systematic changes after certain time points. 
For example, financial time-series data changed abruptly after Asian financial crisis and US housing crisis. 

The most typical model reflecting this type of systematic change is the TAR model. 
The concept of this model can be said to be first conceived when in 1953, P. A. P. Moran raised the problem unsolved 
by linear-type models while he modeled the ecosystem data of Canadian lynx. 
As a solution to this problem in 1983, Tong mentioned the limit of the previous research methods which analyzed the 
time-series data in one frame and proved that it was better to see the time-series data as the combination of various linear 
models with different ranges. He came up with the following model for the Canadian Lynx Ecosystem.
\begin{eqnarray*}
&& x_t = \left\{ 
\begin{array}{ll}
		0.62+1.25x_{t-1}-0.43x_{t-2}+\varepsilon_t, & x_{t-2} \leq 3.25, \\
		2.25+1.52x_{t-1}-1.24x_{t-2}+\varepsilon'_t, & x_{t-2} > 3.25, 
\end{array} \right. \\
&& \varepsilon_t \sim N(0, 0.2^2), ~~ \varepsilon'_t \sim N(0, 0.25^2).
\end{eqnarray*}  
At the same time, he also showed that if the original data of number of sunspots $(\omega_t)$ from 1,749 to 1,924 is 
transformed using Box-Cox transformation, or $x_t = 2( \omega_t^{1/2}-1)$  and then it can be described by the following model:
\begin{eqnarray*}
x_t = \left\{ 
\begin{array}{cc}
		1.9191+0.8416x_{t-1}+0.0728x_{t-2}-0.3153x_{t-3} \\
		+0.1479x_{t-4}-1.985x_{t-5}-0.0005x_{t-6}+0.1875x_{t-7} \\ 
		-0.2701x_{t-8}+0.2116x_{t-9}+0.0091x_{t-10} \\
		+0.0873x_{t-11}+\varepsilon_t, & x_{t-8} \leq 11.9824 \\ 
		4.2746+1.4431x_{t-1}-0.8408x_{t-2} \\
		+0.0554x_{t-3}+\varepsilon_t, & x_{t-8} > 11.9824
\end{array} \right. 
\end{eqnarray*}   
This is a great contribution to analysis of data accompanied by systematic changes.

As shown in the model, Threshold-AR model has been changed to be based on threshold values (11.9824 in the above model) 
with certain time delays (9 in the above model) and became the origin of non-linear time series models by completely 
removing previous linearity.
Therefore, we thought that if we are to model the volatility data with completely different behaviors after certain events 
such as financial crisis, it was better to combine TAR and AARCH attribute within the same structure.

In this paper, we propose TAR-AARCH (Threshold Autoregressive-Asymmetric Autoregressive Conditional Heteroskedasticity) 
Model \eqref{eq1}-\eqref{eq3} to describe volatility.

%\begin{equation*}
%\left\{ 
%\begin{alignedat}{2}
%		& x_t = \sum_{j=1}^t \left(\phi_{j_0}+\sum_{k=1}^p \phi_{jk}x_{t-k}\right) 1 (x_{t-d} \in R_j)+\varepsilon_t & \qquad (1)\\
%		& \varepsilon_t = z_t h_t^{\frac{1}{2}}, z_t : i,i,d, N(0, 1) & \qquad (2) \\
%		& h_t = \alpha_0 + \sum_{i=1}^q (\alpha_i |\varepsilon_{t-i}|+\beta_i \varepsilon^{t-i})^2 & \qquad (3)
%\end{alignedat} 
%\right. 
%\end{equation*}  

\begin{align}
	x_t & =  \sum_{j=1}^t \left(\varphi_{j_0}+\sum_{k=1}^p \varphi_{jk}x_{t-k}\right) 1 (x_{t-d} \in R_j)+\varepsilon_t  \label{eq1}\\
	\varepsilon_t & =  z_t h_t^{\frac{1}{2}}, z_t : i,i,d, N(0, 1) \label{eq2} \\
	h_t & =  \alpha_0 + \sum_{i=1}^q (\alpha_i |\varepsilon_{t-i}|+\beta_i \varepsilon_{t-i})^2 \label{eq3}
\end{align} 
\begin{equation*}
\alpha_0>0, ~ q>0, ~ p, q: \textnormal{known}
\end{equation*}
Here for $t_1, t_2, \cdots, t_l$, the sequence of intervals $R_j, j=1, 2, \cdots$  are as follows:
\begin{equation*}
R_1 = (-\infty, t_1], ~  R_2 = (t_1, t_2], ~ R_3 = (t_2, t_3], \cdots, R_l = (t_{l-1}, +\infty), 
\end{equation*}
\begin{equation*}
1(x \in A) =\left\{
\begin{array}{ll}
1, & x \in A, \\
0, & x \notin A.
\end{array} \right.
\end{equation*} 
 
As shown in the model \eqref{eq1}-\eqref{eq3}, \eqref{eq1} is the TAR model and its error-terms \eqref{eq2} 
and \eqref{eq3} are AARCH models. 
In other words, model \eqref{eq1}-\eqref{eq3} form the TAR model is inclusive of Asymmetric ARCH effect. 

     Also given the consideration that complete form of likelihood function is impossible in the model, 
the method of appropriate parameter estimation based on QMLE (Quasi Maximum Likelihood Estimation) has been established 
and asymptotic normality of estimators have been proved for this TAR-AARCH model. 
Then, it is possible to estimate all the parameters of the combined TAR-AARCH model after estimation of time delay 
and threshold parameter through wavelet in the TAR model \cite{kim2}.

%
% --------------  2  Estimation of parameters in the TAR-ARCH model  -----------------
%
	
\section{Estimation of parameters in the TAR-ARCH model}

To use QMLE for the model, we should first find $\frac{\partial h_t}{\partial \alpha_t}, \frac{\partial h_t}{\partial \varphi_{jk}}, \frac{\partial \varepsilon_t}{\partial \alpha_i}, \frac{\partial \varepsilon_t}{\partial \varphi_{jk}}$. 
But as the \eqref{eq3} shows, it is impossible to estimate $\frac{\partial h_t (\theta)}{\partial \varphi_{jk}}$  as the 
parameter  $\varphi_{jk}$ contains absolute value term. 
Therefore usual QMLE method will be ineffective in finding out QMLE for parameter.

But if concentrated QMLE is to be used for \eqref{eq1}-\eqref{eq3}, this problem can be solved. 
That is, $\bar{\alpha} \overset{\Delta}{=} (\alpha_0, \cdots, \alpha_q, \beta_1, \cdots, \beta_q)$  can be fixed and 
QMLE concentrated on $\bar{\theta}_1 \overset{\Delta}{=} (\varphi_{10}, \cdots, \varphi_{lp})$  can be obtained for \eqref{eq1} 
and its asymptotic normality can be observed.

If we assume that $\bar{\theta}_1 = (\varphi_{10}, \cdots, \varphi_{lp})$  is known, obtain QMLE concentrated on $\bar{\alpha}$  for $x_t$  and demonstrate 
its asymptotic normality, then we will be able to obtain estimators for parameters $\bar{\theta}_1$  and  $\bar{\alpha}$ 
through the two steps above.
But we should be able to ascertain if such estimators can be assumed to be the estimators for parameters, and if so, 
how much difference they have when compared to QMLE. 
For this, the parameters of the TAR-ARCH model with QMLE and its asymptotic normality already obtained have been divided 
as based of the above mentioned method to obtain concentrated QMLE and asymptotic normality for each sub-parameter 
and the results have been compared with QMLE obtained from \cite{kim1}, to prove efficiency of this method.
%
% --------------  2.1  Concentrated QMLE in the TAR-ARCH model  -----------------
%
\subsection{Concentrated QMLE in the TAR-ARCH model}
Let
\begin{align*}
& \alpha \overset{\Delta}{=} (\alpha_0, \alpha_1, \cdots, \alpha_q), \\
& \theta_1 \overset{\Delta}{=} (\varphi_{10}, \varphi_{11}, \cdots, \varphi_{1p}, \varphi_{20}, \varphi_{21}, \cdots, \varphi_{2p}, \cdots, \varphi_{l0}, \varphi_{l1}, \cdots, \varphi_{lp}).
\end{align*}
If for TAR-ARCH model, $\alpha$  is known, then QML equation concentrated on $\theta_1$  is formulated as:
\begin{equation*}
\left\{ 
\begin{aligned}
	\sum_{t=1}^n \frac{1}{h_t(\alpha)} \varepsilon_t(\theta_1){\bf 1}(x_{t-d} \in {\bf R}_j) = 0, \\
	\sum_{t=1}^n \frac{\varepsilon_t(\theta_1)}{h_t(\alpha)} x_{t-k}{\bf 1}(x_{t-d} \in {\bf R}_j) = 0,
\end{aligned} 
\right. 
\end{equation*}  
\begin{equation*}
j=1, \cdots, l, ~~ k=1, \cdots, p
\end{equation*}  

%
%%%%%%%%%%%%%   Theorem 1   %%%%%%%%%%
%

\begin{thr} \label{thr1}
In the model \eqref{eq1}-\eqref{eq3}, let $\hat{\theta}_{1, n}$  be QML estimator concentrated on $\theta_1$  
when $\alpha$  is known, and also that it satisfies the strong stationary condition
\begin{equation*}
\theta_1 \in \boldsymbol{\Theta}, ~~ \alpha \in \boldsymbol{\Theta}, ~~  \hat{\theta}_{1, n} \in \boldsymbol{\Theta}.
\end{equation*}  
$\theta_{1, 0} \in \boldsymbol{\Theta}$  is a true parameter of model \eqref{eq1}-\eqref{eq3} when $\alpha$ is known, then
\begin{enumerate}
\item[$1^{\circ}$]  $\hat{\theta}_{1, n} \overset{p}{\to} \theta_{1, 0}$, 

\item[$2^{\circ}$] $\sqrt{n}\left( \hat{\theta}_{1, n}-\theta_{1,0} \right) \sim N\left( 0, {\bf I}^{-1}(\theta_{1, 0}) \right)$.
\end{enumerate}
\end{thr}
Likewise, QMLE concentrated on   and its asymptotic normality can be proven in the same way as Theorem \ref{thr1}.

We can see how much difference the QMLE and the concentrated QMLE have as follows.

For this, parameter obtained with QMLE (TAR-ARCH) may be divided into $\left( \tilde{\theta}_1, \tilde{\alpha}\right)$ 
and its Fisher’s Information matrix $I^{-1}(\theta) = I^{-1}\left( \tilde{\theta}_1, \tilde{\alpha}\right)$  can be 
compared to the Fisher’s Information matrix of the above concentrated QMLE,  $I^{-1}(\theta_1)$ $= I^{-1}(\alpha)$ to prove the following relation between them:
\begin{align*}
& \left[ \textnormal{var} \left( \sqrt{n}\left(\tilde{\theta}_{1, n}-\theta_{1, 0}\right) \right)\right]^{-1} \geq  \left[ \textnormal{var} \left( \sqrt{n}\left(\hat{\theta}_{1, n}-\theta_{1, 0}\right) \right)\right]^{-1}, \\
& \left[ \textnormal{var} \left( \sqrt{n}\left(\hat{\alpha}_{n}-\alpha_{0}\right) \right)\right]^{-1} = \left[ \textnormal{var} \left( \sqrt{n}\left(\tilde{\alpha}_{n}-\alpha_{0}\right) \right)\right]^{-1}.
\end{align*} 
\begin{equation*}
( \tilde{\theta}_{1, n}, \tilde{\alpha}_n : \textnormal{sub-parameters of QMLE})
\end{equation*} 
Therefore it is concluded that if the above method for sub-parameters $\theta_1$, $\alpha$  is used to obtain concentrated QMLE for the TAR-AARCH models which are difficult to apply QMLE, the obtained estimators will be acceptable although the efficiency is a little diminished.

%
% --------------  2.2  Asymptotic normality of the concentrated QMLE in the TAR-AARCH model  -----------------
%
\subsection{Asymptotic normality of the concentrated QMLE in the TAR-AARCH model}
       
As easily seen, in the model \eqref{eq1}-\eqref{eq3}, if  $\bar{\alpha}$ is known, the QML equation concentrated 
on $\bar{\theta}_1$  is as follows:
\begin{equation*}
\left\{ 
\begin{aligned}
	\sum_{t=1}^n \frac{1}{h_t(\bar{\alpha})} \varepsilon_t(\bar{\theta}_1){\bf 1}(x_{t-d} \in {\bf R}_j) = 0, \\
	\sum_{t=1}^n \frac{\varepsilon_t(\bar{\theta}_1)}{h_t(\bar{\alpha})} x_{t-k}{\bf 1}(x_{t-d} \in {\bf R}_j) = 0,
\end{aligned} 
\right. 
\end{equation*}  
\begin{equation*}
j=1, \cdots, l, ~~ k=1, \cdots, p
\end{equation*}   
Then, it is possible to prove the following theorem.
%
%%%%%%%%%%%%%   Theorem 2   %%%%%%%%%%
%

\begin{thr} \label{thr2}
In the model \eqref{eq1}-\eqref{eq3}, $\bar{\alpha}$  is known and $\bar{\theta}_{1, n}$  is the QML estimator concentrated on $\bar{\theta}_1$. 
Also, it satisfies the strong stationary condition for $\bar{\theta}_1 \in \boldsymbol{\Theta}, ~~ \bar{\theta}_{1, n}, \bar{\alpha} \in \boldsymbol{\Theta}$. 
If $\bar{\theta}_{1, 0}$ is the true parameter for model \eqref{eq1}-\eqref{eq3} when $\bar{\alpha}$ is known, then the following is true:
\begin{enumerate}
\item[$1^{\circ}$]  $\bar{\theta}_{1, n} \overset{p}{\to} \bar{\theta}_{1, 0}$, 

\item[$2^{\circ}$] $\sqrt{n}\left( \bar{\theta}_{1, n}-\bar{\theta}_{1,0} \right) \sim N\left( 0, {\bf I}^{-1}\left(\bar{\theta}_{1, 0}\right) \right)$.
\end{enumerate} 
The QMLE concentrated on $\bar{\alpha}$  can be obtained and its asymptotic normality proven in the same way.
\end{thr}

%
% --------------  3  Conclusion  -----------------
%
	
\section{Conclusion}

In this paper, first, two types of non-linear models, TAR and Asymmetric ARCH were combined into TAR-AARCH model to describe volatility in a more effective way. Second, appropriate parameter estimation method was proposed to counter various difficulties generated by asymmetry of the combined model.


\begin{thebibliography}{20}

\bibitem{bla}
F. Black, M. Scholes, The pricing of options and corporate liabilities, Journal of Political Economy, 81(3) (1973), 637-654.

\bibitem{bol}
T. Bollerslev, Generalized autoregressive conditional heteroskedasticity, Journal of Econometrics, 31(3) (1986), 307-327.

\bibitem{eng}
R.F. Engle, Risk and volatility: Economic models and financial practice, Nobel lecture, Stockholm, December 8, 29, 2003.

\bibitem{glo}
L.R. Glosten, R. Jagannathan, D.E. Runkle, On the relation between the expected value and the volatility of the nominal excess return on stocks, Journal of Finance, 48(5) (1993), 1779-1801.

\bibitem{har}
W. H\"ardle, T. Kleinow, G. Stahl, Applied quantitative finance: Theory and computational tools, Springer, 2002.

\bibitem{kim1}
S.Y. Kim, Threshold AR with Asymmetric ARCH type Error (in Korean), Dissertation for ph D, 2007.

\bibitem{kim2}
S.Y. Kim, M.C. Kim, The identification of thresholds and time delay in self-exciting TAR model by wavelet, International Symposium in Commemoration of the 65th Anniversary of the Foundation of \textbf{Kim Il Sung} University (Mathematics), 20-21. Sep. Juche100 (2011)  Pyongyang, DPR Korea,  arXiv 1303.4867 [math-ph].

\bibitem{nel}
D.B. Nelson, Conditional heteroskedasticity in assert returns: a new approach, Econometrica, 59(2) (1991), 347-370.

\bibitem{ton}
H. Tong, Non-linear time series: A dynamical system approach, Clarendon Press, Oxford, 1990.

\bibitem{liu}
%\begin{CJK*}{UTF8}{zhsong}
%刘继春, 一个非对成GARCH模型的严平稳遍历性, 厦门大学学报 (自然科学版), 42(2) (2003), 153-156.
%\clearpage\end{CJK*}
\begin{CJK*}{UTF8}{gbsn}
刘继春, 一个非对成GARCH模型的严平稳遍历性, 厦门大学学报 (自然科学版), 42(2) (2003), 153-156.
\clearpage\end{CJK*}

\end{thebibliography}
 	\end{document}